\documentclass[
prd,aps,twocolumn, showpacs,tightenlines,nofootinbib,preprintnumbers, superscriptaddress
]{revtex4}
\usepackage{graphicx}
\usepackage{amssymb,amsmath,latexsym}
\usepackage{amsfonts}

\usepackage{cancel}
\usepackage{color}
\input{colordvi.tex}
\input{epsf}
\usepackage{epsf}
\usepackage{graphicx,epsfig}
\usepackage{bm}
\usepackage{latexsym,float}

\newcommand{\bear}{\begin{array}}  \newcommand{\eear}{\end{array}}
\newcommand{\beq}{\begin{equation}}  \newcommand{\eeq}{\end{equation}}
\newcommand{\bef}{\begin{figure}}  \newcommand{\eef}{\end{figure}}
\newcommand{\bec}{\begin{center}}  \newcommand{\eec}{\end{center}}

\newcommand{\beqa}{\begin{eqnarray}}
\newcommand{\eeqa}{\end{eqnarray}}

\newcommand {\ga} {\ {\raise-.5ex\hbox{$\buildrel>\over\sim$}}\ }
\newcommand {\la} {\ {\raise-.5ex\hbox{$\buildrel<\over\sim$}}\ } 

\newcommand{\fig}[1] {Fig.~\ref{#1}}

\def\be{\begin{equation}}
\def\ee{\end{equation}}
\def\ba{\begin{eqnarray}}
\def\ea{\end{eqnarray}}
\renewcommand{\(}{\left(} 
\renewcommand{\)}{\right)} 
\renewcommand{\[}{\left[} 
\renewcommand{\]}{\right]} 

\begin{document}

\title{Slow-roll freezing quintessence}

\author{Sourish Dutta}
\email{sourish.d@gmail.com}
\affiliation{Department of Physics and School of Earth and Space Exploration,
Arizona State University, Tempe, AZ 85287}

\author{Robert~J. Scherrer}
\email{robert.scherrer@vanderbilt.edu}
\affiliation{Department of Physics and Astronomy,
Vanderbilt University,
Nashville, TN 37235}

\date{\today}

\pacs{98.80.Cq ; 95.36.+x }

\begin{abstract}
We examine the evolution of quintessence models with potentials
satisfying $(V^\prime/V)^2 \ll 1$ and $V^{\prime \prime}/V \ll 1$,
in the case where the initial field velocity is nonzero.  We derive an
analytic approximation for the evolution of the equation of state parameter,
$w$, for the quintessence field.  We show
that such models
are characterized by an initial rapid freezing phase, in which
the equation of state parameter $w$ decreases with time, followed by slow
thawing evolution, for which $w$ increases with time.  
These models resemble constant-$V$ models at early times but diverge
at late times.
Our analytic approximation gives results in excellent agreement
with exact numerical evolution.
\end{abstract}

\maketitle

\section{Introduction}

Cosmological data from a wide range of sources including type Ia supernovae \cite{union08, perivol, hicken}, the cosmic microwave background \cite{Komatsu}, baryon acoustic oscillations \cite{bao,percival}, cluster gas fractions \cite{Samushia2007,Ettori} and gamma ray bursts \cite{Wang,Samushia2009} seem to indicate that at 
least 70\% of the energy density in the
universe is in the form of an exotic, negative-pressure component,
called dark energy.  

The dark energy component is usefully parameterized by its equation of state parameter, 
defined as the ratio of its pressure to its density:
\be
\label{w}
w=p_{\rm DE}/\rho_{\rm DE}.
\ee
Observations constrain $w$ to be very close to $-1$.
For example, if $w$ is assumed to be constant, then $-1.1 \la w \la -0.9$  \cite{Wood-Vasey,Davis}.

While a cosmological constant ($w = -1$) remains consistent with the
observations,
a variety of other models have been proposed in which $w$ is time varying.
A common approach is to use a scalar field $\phi$ as the dark energy component.
The class of models in which the scalar field is canonical is dubbed quintessence
\cite{RatraPeebles,CaldwellDaveSteinhardt,LiddleScherrer,SteinhardtWangZlatev}
and has been extensively studied. (See Ref. \cite{Copeland} for a recent
review). 

A related, yet somewhat different approach is phantom dark energy, i.e., a component for which 
$w<-1$, as first proposed by Caldwell \cite{Caldwell}. Such models have well-known problems
\cite{CarrollHoffmanTrodden,ClineJeonMoore,BuniyHsu,BuniyHsuMurray} (however see 
\cite{Creminelli:2008wc,Cai} for recent attempts to construct a stable model),
but nevertheless have been widely studied as potential dark energy candidates.

These models allow considerable freedom in the choice of the potential $V(\phi)$, leading
to an infinite set of possible models and corresponding behaviors for the evolution $w$
as a function of the redshift $z$.
It would therefore be a considerable simplification
if one could find an interesting subset of such
models that converged to a single trajectory or a well-defined set of trajectories for $w(z)$.
One approach that yields such a simplification was considered by
Scherrer and Sen \cite{ScherrerSen1},
who examined the evolution of a scalar field in a ``nearly flat" potential, where
the flatness condition consisted of the slow-roll conditions familiar from inflation:
\ba
\label{SR1}\lambda^2 \equiv \(- \frac{1}{V}\frac{dV}{d\phi}\)^2\ll 1,\\
\label{SR2}\left|\frac{1}{V}\frac{d^2 V}{d\phi^2}\right|\ll 1.
\ea
As in Ref. \cite{ScherrerSen1}, we will assume for definiteness that
$dV/d\phi < 0$ and $\lambda > 0$, but of course none our results will
depend on this choice.  Note that while equations (\ref{SR1}) and (\ref{SR2})
are the familiar slow-roll conditions from inflation, the evolution
of the scalar field is very different for the case of quintessence,
since in that case one must also include the effect of the
matter density on the expansion rate.

Ref. \cite{ScherrerSen1} considered models satisfying equations
(\ref{SR1}) and (\ref{SR2}) in
which the field is initially at rest, so that $w \approx -1$ at early times, and
$w$ increases at late times as the field rolls down the potential;
in the terminology of Ref. \cite{CL}, these are ``thawing" models.
Then equation (\ref{SR1}) ensures that $w$ remains close to $-1$, while
equations (\ref{SR1}) and (\ref{SR2}) taken together indicate that $\lambda$ is nearly
constant.  
For all potentials satisfying these conditions,
it can be shown that the
behavior of $w$ can be accurately described by
a unique function of $\Omega_\phi$ (the fraction of the total
density contributed by the quintessence field, where we assume a flat universe)
and the (assumed constant) value of $\lambda$ \cite{ScherrerSen1}.
In \cite{ScherrerSen2} this result was extended to
phantom models satisfying
Eqs.~(\ref{SR1}-\ref{SR2}). 

In this paper, we relax the initial condition
that $\dot \phi \equiv d\phi/dt = 0$,
so that $w \ne -1$ initially, but we retain both
slow roll conditions on the potential (equations \ref{SR1}
and \ref{SR2}).  This allows us to extend the
formalism of Ref. \cite{ScherrerSen1} to models in which $w$ decreases
toward $-1$ at late times, rather than increasing away from $-1$; the
former models are called ``freezing" models \cite{CL}.

Note that the slow roll conditions, Eqs.~(\ref{SR1}-\ref{SR2}),
while sufficient to ensure $w\approx -1$ today, are not necessary,
and many other attempts to classify or simplify the set of quintessence
trajectories have been proposed.  For example, if equation
(\ref{SR1}) holds, but equation (\ref{SR2}) is relaxed,
one still has $w \approx -1$, but there is now an
extra degree of freedom, the value of $V^{\prime \prime}/V$.
Instead of a single solution for the evolution of $w$,
one obtains a well-defined family of solutions \cite{ds1}.
(This family of solutions includes
the slow-roll solution of Ref. \cite{ScherrerSen1} as a
special case in the limit
where $V^{\prime \prime}/V \rightarrow 0$). These
models were explored in more detail
in Refs. \cite{ds2,ds3,chiba}.

Other attempts to systematize the behavior of dark-energy quintessence fields based
on other assumptions have been
given in Refs. \cite{Watson,Crit,Neupane,Cahn,Chiba,Cortes,Bond,Luo}.  Where
these other approaches overlap those of this paper, further discussion
and comparison with our results will be given below.

In the next section, we examine the evolution of $w$ for slow-roll potentials with
the assumption that $\dot \phi \ne 0$ initially.  In Sec. III, we compare our predictions
for $w(a)$ with exact numerical results.
Our conclusions are discussed in Sec. IV.

\section{Evolution of $w$: Analytic results}

We will assume that the dark energy is provided by a minimally-coupled
scalar field, $\phi$, with equation of motion given by
\begin{equation}
\label{motionq}
\ddot{\phi}+ 3H\dot{\phi} + \frac{dV}{d\phi} =0,
\end{equation}
where $H$ is the Hubble parameter, given by
\begin{equation}
\label{H}
H = \left(\frac{\dot{a}}{a}\right) = \sqrt{\rho/3}.
\end{equation}
Here $a$ is the scale factor, $\rho$ is the total density, and we
take $8 \pi G = 1$ throughout.
Equation (\ref{motionq}) indicates
that the field rolls downhill in the potential $V(\phi)$,
but its motion is damped by a term proportional to $H$.

The pressure and density of the
scalar field are given by
\begin{equation}
p = \frac{\dot \phi^2}{2} - V(\phi),
\end{equation}
and
\begin{equation}
\label{rhodense}
\rho = \frac{\dot \phi^2}{2} + V(\phi),
\end{equation}
respectively, and the equation of state parameter, $w$, for the quintessence
field
is given by equation (\ref{w}).

As in Ref. \cite{ScherrerSen1}, we note that the equations simplify
if we express $w$ as a function of $\Omega_\phi$, giving
\begin{equation}
\label{exact}
\frac{dw}{d\Omega_\phi}
= \frac{-3(1+w)(1-w) + \lambda(1-w)\sqrt{3 (1+w) \Omega_\phi}}
{3(-w)\Omega_\phi(1-\Omega_\phi)}.
\end{equation}
(See Ref. \cite{ScherrerSen1} for the details of this derivation).

Consider a scalar field moving in a potential that satisfies equations
(\ref{SR1}) and (\ref{SR2}), and assume we are in a regime
such that $1+w \ll 1$.  Note that in relaxing the initial condition that
$\dot \phi = 0$, it is possible to have transient evolution with
$w$ far from $-1$ even when equations (\ref{SR1})
and (\ref{SR2}) are satisfied.  Here we assume that the evolution
has proceeded far enough that
$1+w \ll 1$, but $w \ne -1$, when we begin to examine the evolution.
With these assumptions, equation (\ref{exact}) simplifies to
\cite{ScherrerSen1}
\begin{equation}
\label{dwdO}
\frac{dw}{d\Omega_\phi} = - \frac{2 (1+w)}{\Omega_\phi (1-\Omega_\phi)}
+ \frac{2}{3} \lambda_0 \frac{\sqrt{3
(1+w)}}{(1-\Omega_\phi)\sqrt{\Omega_\phi}},
\end{equation}
where $\lambda_0$ is the (assumed constant) value of $\lambda$.
The general solution of equation (\ref{dwdO}) is
\begin{align}
\label{gensol}
&1+w=\frac13\lambda_0^2 \nonumber\\
&\times\[\frac{1}{\sqrt{\Omega_\phi}}-\(\frac{1}{\Omega_\phi}-1\)
\(\tanh^{-1}\(\sqrt{\Omega_\phi}\)+C\)\]^2.
\end{align}

The constant $C$ parametrizes different initial values of $\dot \phi$, and
therefore characterizes different evolutionary trajectories.  Ref. \cite{ScherrerSen1}
considered only the case $C=0$, which corresponds to
$\dot \phi = 0$ initially, giving
\begin{align}
\label{C0}
&1+w=\frac13\lambda_0^2 \nonumber\\
&\times\[\frac{1}{\sqrt{\Omega_\phi}}-\(\frac{1}{\Omega_\phi}-1\)\tanh^{-1}\sqrt{\Omega_\phi}\]^2.
\end{align}
Equation (\ref{C0}), derived as one of the main results
in Ref. \cite{ScherrerSen1}, displays a purely thawing behavior,
as the field begins initially with $w = -1$, and $w$ increases
as $\phi$ rolls down the potential.  Note that at early times, when $\Omega_\phi
\ll 1$,
we can expand equation (\ref{C0}) to give\
\begin{equation}
1+w = \frac{1}{3} \lambda_0^2 \[\frac{2}{3} \Omega^{1/2} + \frac{2}{15} \Omega^{3/2}
+ \frac{2}{35} \Omega^{5/2}+...\]^2.
\end{equation}
Retaining only the first term in this expansion gives 
\begin{equation}
1+w = \frac{4}{27} \lambda_0^2 \Omega_\phi.
\end{equation}
This agrees with the result of Cahn, de Putter, and Linder \cite{Cahn}, who defined
a ``flow parameter" $F \equiv (1+w)/\lambda_0^2 \Omega_\phi$ and showed that $F = 4/27$ in
the limit considered here.

Equation (\ref{gensol}) when $C \ne 0$ corresponds to the case where the field has $\dot \phi \ne 0$
initially.  The case $C < 0$ corresponds to a field rolling down the potential, while $C > 0$
gives solutions for the field rolling uphill initially.  While the latter possibility might
seem somewhat contrived, it has been considered previously \cite{Csaki,Sahlen1,Sahlen2} and we include it here for
completeness.

The actual value of $C$ depends in the initial value of $\dot \phi$, but it is
easiest to evaluate $C$
in terms of the value of $w$ corresponding to some initial value of $\Omega_\phi$.  If we designate
these values as $w_i$ and $\Omega_{\phi i}$, respectively, and take an early enough epoch that
$\Omega_{\phi i} \ll 1$, then we have
\begin{equation}
C = \pm \frac{\sqrt{3(1+w_i)}\Omega_{\phi i}}{\lambda_0},
\end{equation}
where the sign of $C$ is determined by whether the field is rolling initially up or down the potential.
In terms of $\dot \phi$, we simply have $1+w_i \approx \dot \phi^2/V_i$, where $V_i$ is the
(initial but assumed nearly constant) value of the potential.

\begin{center}
\begin{figure*}
\begin{tabular}{c@{\qquad}c}
\epsfig{file=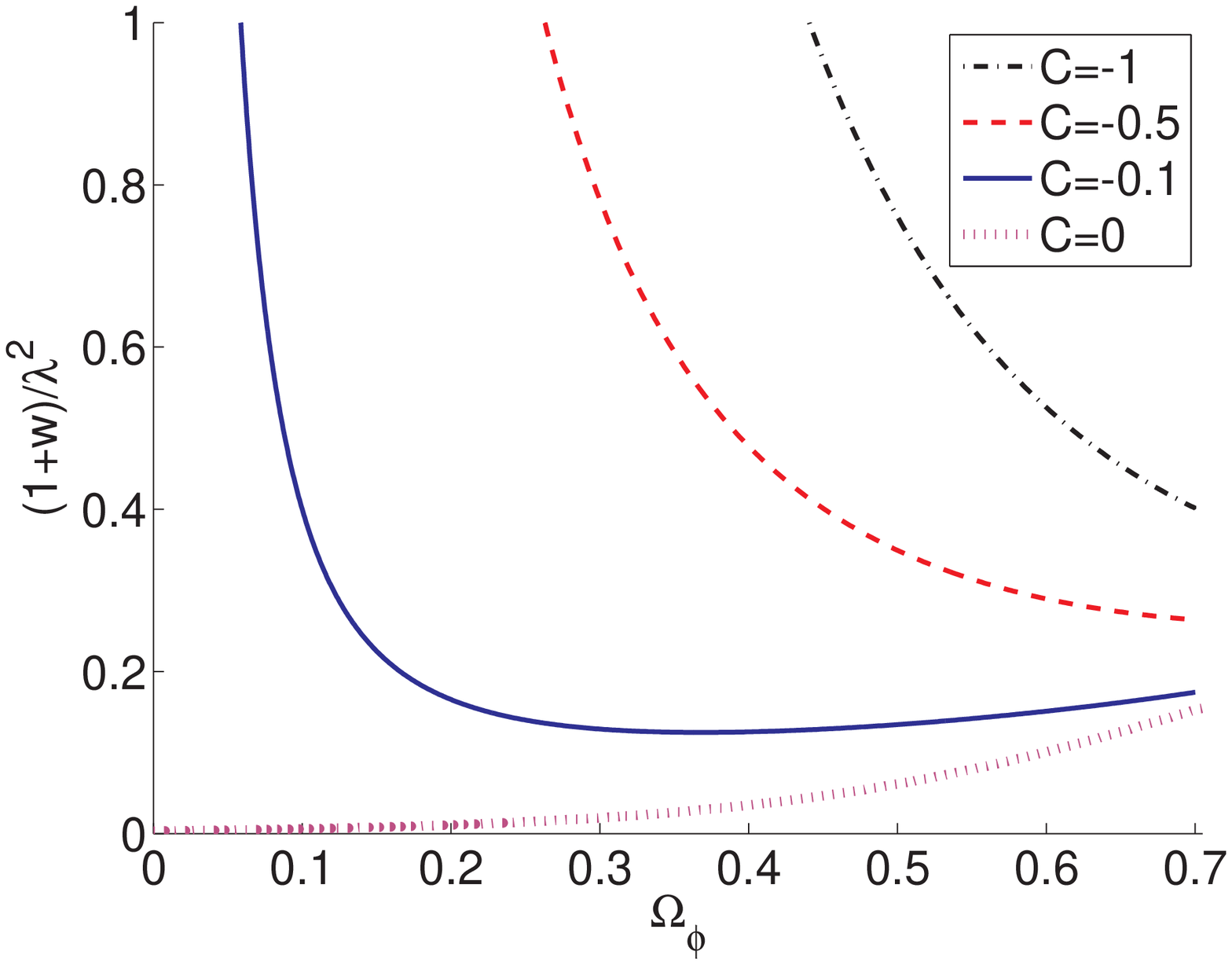,width=7 cm}&\epsfig{file=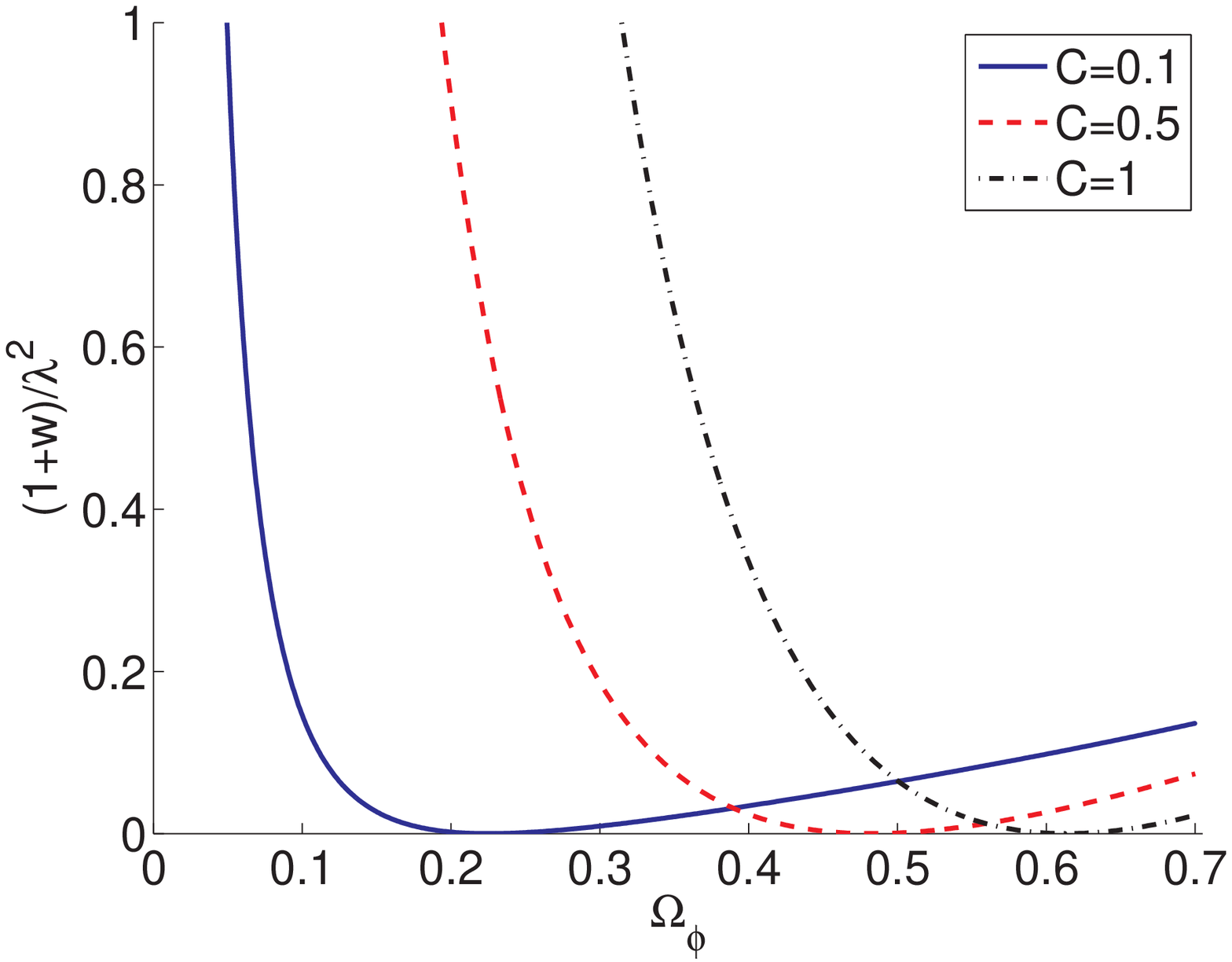,width=7 cm}
\end{tabular}
\caption{\label{1plusw_Om}
The value of $(1+w)/\lambda_0^2$ as a function of $\Omega_\phi$ for the
indicated values of C.  Here $w$ is the equation of state parameter for the quintessence field,
$\Omega_\phi$ is the fraction of the total density contributed by the quintessence field, and
$\lambda_0$ is the
value of $-(1/V)(dV/d\phi)$ (taken to be constant).}
\end{figure*}
\end{center}

In equation (\ref{gensol}) $w$ enters only
in the combination $(1+w)/\lambda_0^2$.  In the left panel of \fig{1plusw_Om}, we display this quantity as a function of
$\Omega_\phi$ for the indicated values of $C \le 0$.  Except for $C = 0$, the generic
behavior of these models is an initial freezing period, during which $w$ decreases rapidly with time,
followed by a period of thawing, with a much slower increase of
$w$ with time.  

The value of $\Omega_\phi$ for which $1+w$ reaches a minimum,
corresponding to the transition between freezing and thawing behavior, can easily be derived
from equation (\ref{gensol}):
\begin{equation}
\label{transition}
C = \sqrt{\Omega_\phi} - \tanh^{-1} \sqrt{\Omega_\phi}.
\end{equation}
Thus, models with $-C \ga 0.5$ are still freezing at the present.  On the other hand,
it is clear from Fig. 1 that
models with $-C \la 0.1$ are nearly indistinguishable from
the $C=0$ case, as they enter the thawing period when $\Omega_\phi \ll 1$.

In the right panel of \fig{1plusw_Om}, we show the corresponding evolution of $(1+w)/\lambda_0^2$ as a function of $\Omega_\phi$ for
$C>0$.  These models show more complex behavior, as the field first rolls uphill, stops (giving $w = -1$),
and then rolls back down again.  The value of $\Omega_\phi$ at which $w = -1$ is given
by
\begin{equation}
C = \frac{\sqrt{\Omega_\phi}}{\(1-\Omega\)} - \tanh^{-1} \sqrt{\Omega_\phi}.
\end{equation}

\begin{center}
\begin{figure*}
\begin{tabular}{c@{\qquad}c}
\epsfig{file=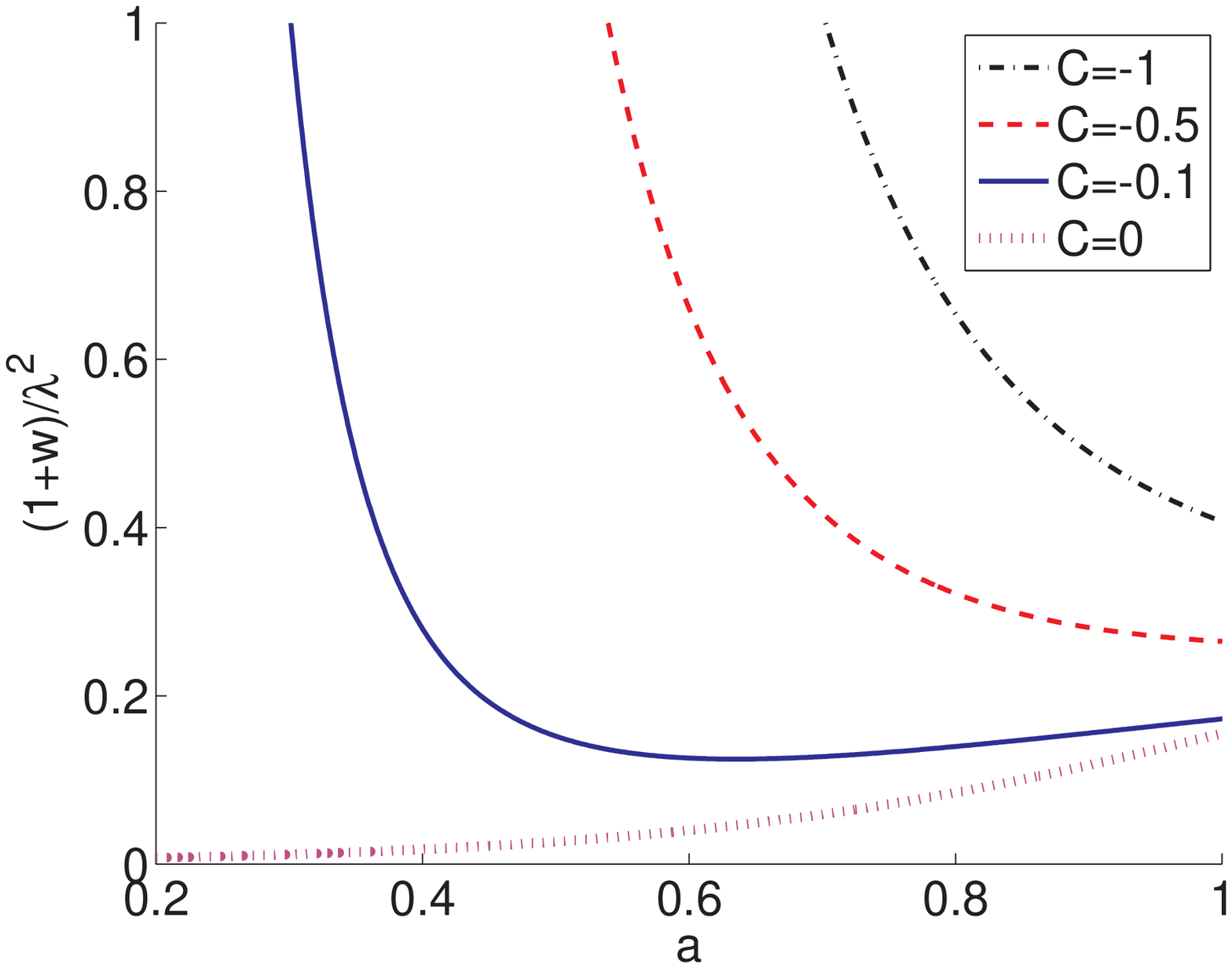,width=7 cm}&\epsfig{file=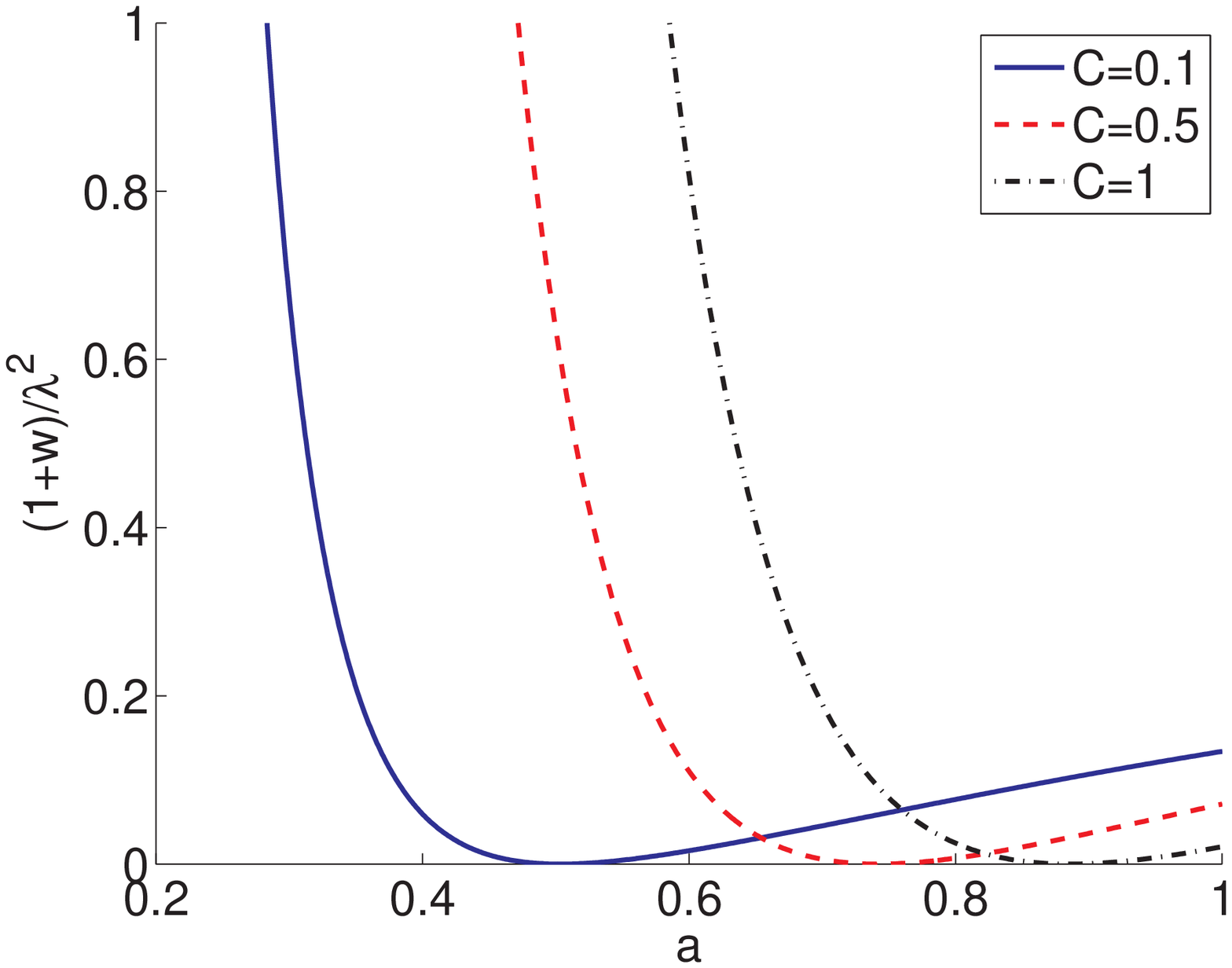,width=7 cm}
\end{tabular}
\caption{\label{1plusw_a}
As \fig{1plusw_Om}, with the scale factor $a$ (taken to be 1 at the present) on the horizontal axis,
and $\Omega_{\phi 0} = 0.7$.}
\end{figure*}
\end{center}

Since we always assume $1+w \ll 1$,
we can express $w$ as a function of the scale factor $a$ by taking
$\Omega_\phi$ to be well-approximated by its value for a $\Lambda$CDM universe \cite{ScherrerSen1}:
\be
\label{Omega(a)}
\Omega_\phi=\[1+\(\Omega_{\phi 0}^{-1}-1\)a^{-3}\]^{-1}.
\ee
Equations (\ref{gensol}) and (\ref{Omega(a)}) together then provide an analytic approximation for
$w(a)$ for these models.  In Fig. \ref{1plusw_a}, we show $(1+w)/\lambda_0^2$ as a function of
the scale factor for the same set of models as in \fig{1plusw_Om},
where we have taken $\Omega_{\phi 0} = 0.7$ and $a=1$ at the present.

%
%

\section{Comparison with exact evolution}

We now test the accuracy of the analytic approximation derived in the previous
section by comparing our analytic predictions for $w(a)$ with
an exact numerical integration of the equations of motion.  Although we expect
our approximation to apply to any potential with a region in $\phi$
satisfying equations (\ref{SR1}) and (\ref{SR2}), we will consider
here two representive examples:

\begin{itemize}
\item{Linear potential}
\be
\label{linearpotential}
V\(\phi\)=V_0-\alpha\phi
\ee
where take $\alpha=0.1$.

\item{Power-Law potential}
\be
V\(\phi\)= V_0 \phi^{-n}
\label{powerlawpotential}
\ee
where we take $n=0.1.$
\end{itemize}
In both of these cases the parameters of the potential were
chosen so as to satisfy
the slow-roll conditions (equations \ref{SR1} and \ref{SR2}),
so that $\lambda$ is roughly constant and $\ll
1$. 
We set the initial value of $w$ to $w_i=-0.95$
and $\lambda_0=-0.08$. Then the initial value of $\Omega_\phi$
is chosen to
give $C=\pm0.1$, $C=\pm0.5$
and $C=\pm1$. (Numerical tests of the $C=0$ case
can be found in Ref. \cite{ScherrerSen1}).

\begin{center}
\begin{figure*}
\begin{tabular}{c@{\qquad}c}
\epsfig{file=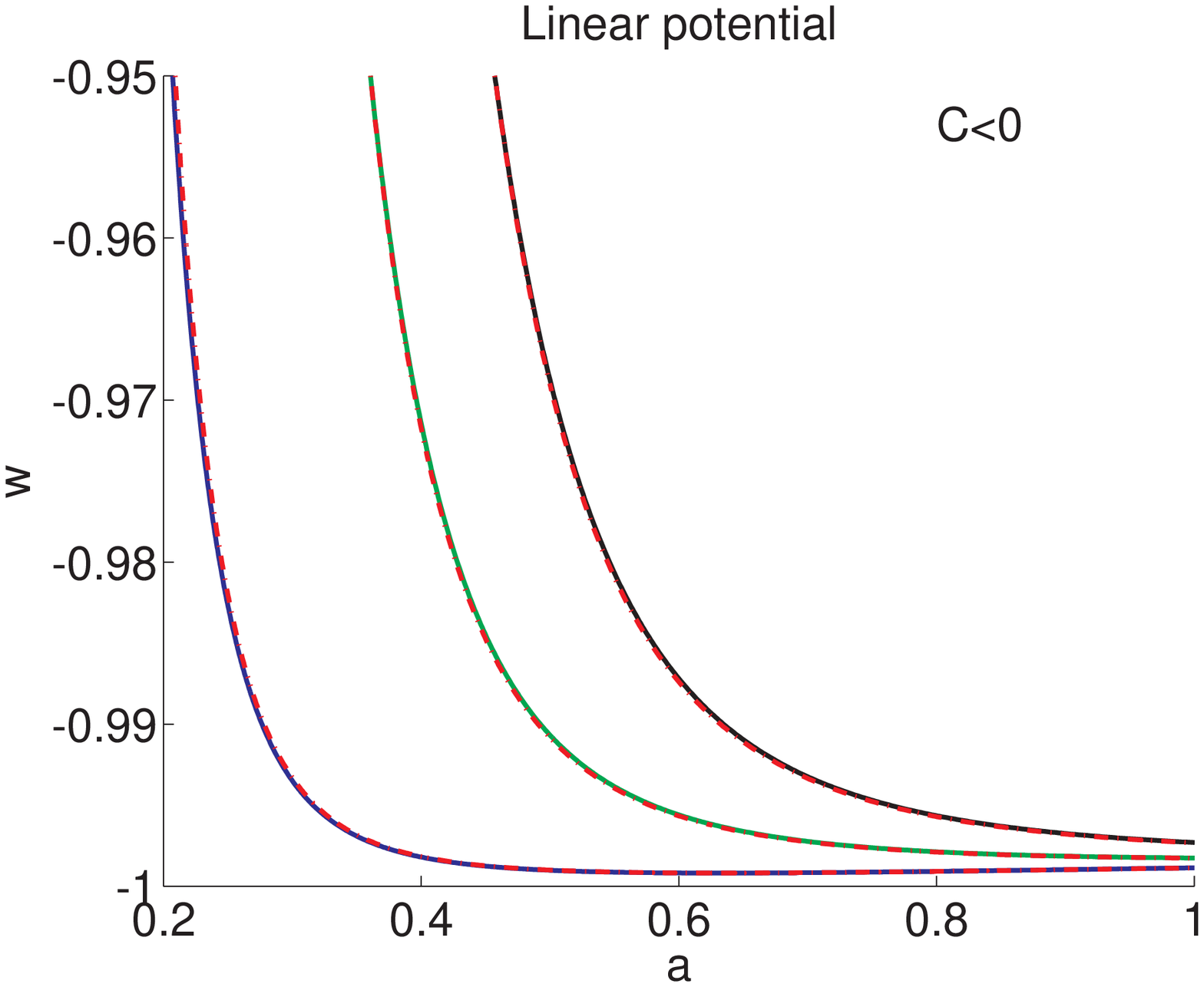,width=7 cm}&\epsfig{file=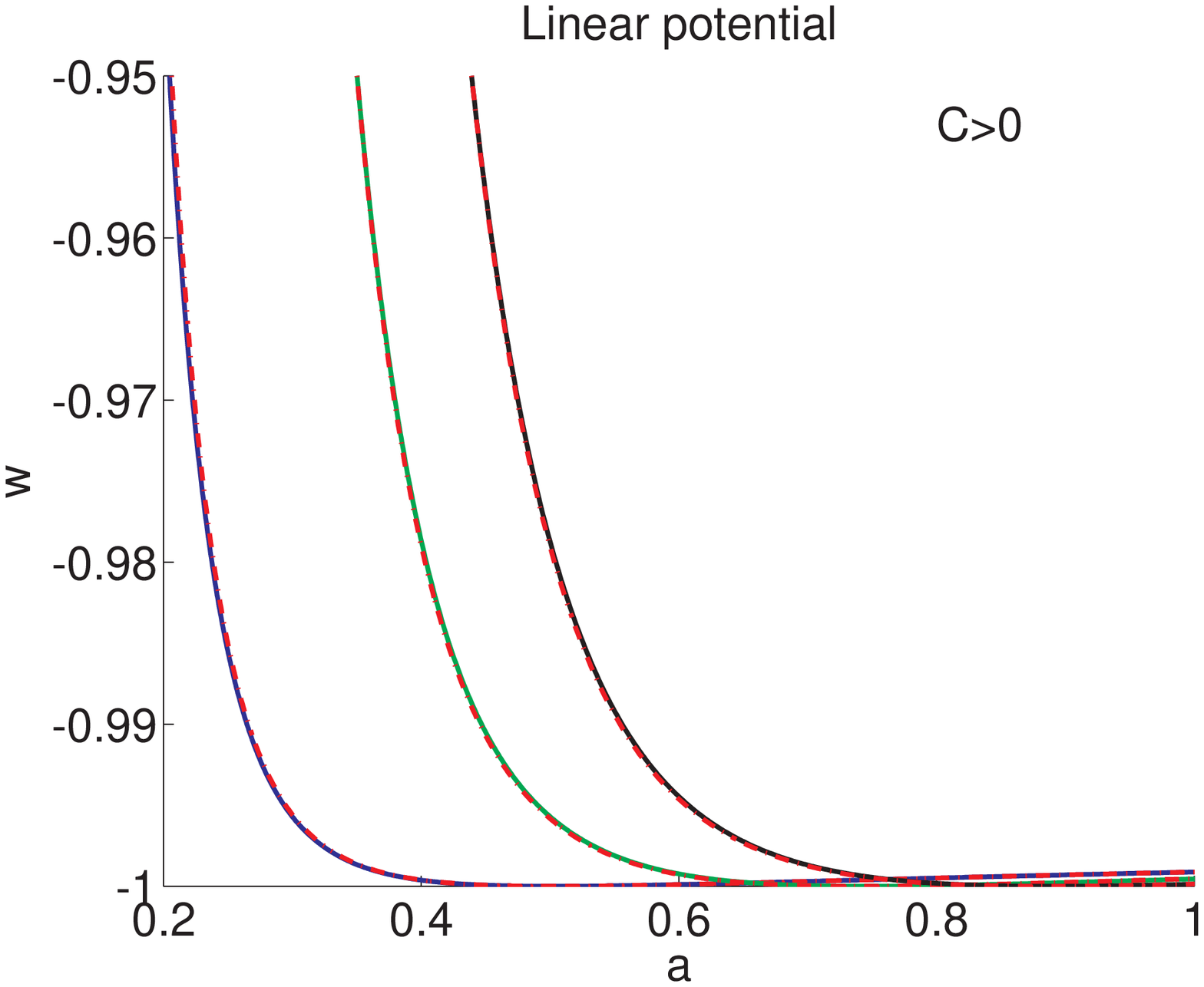,width=7 cm}
\end{tabular}
\caption{\label{w_a_linear}
Numerical versus analytic evolution of the equation of state parameter, $w$, for the linear potential. From left
to right the plots correspond to $|C|=0.1$, $|C|= 0.5$ and $|C|=1$ respectively.
The sign of $C$ is indicated on the figure.
Solid curves correspond to the exact (numerically-derived) evolution,
while the dashed curves indicate the analytical approximation from equations
(\ref{gensol}) and (\ref{Omega(a)}).}
\end{figure*}
\end{center}

\begin{center}
\begin{figure*}
\begin{tabular}{c@{\qquad}c}
\epsfig{file=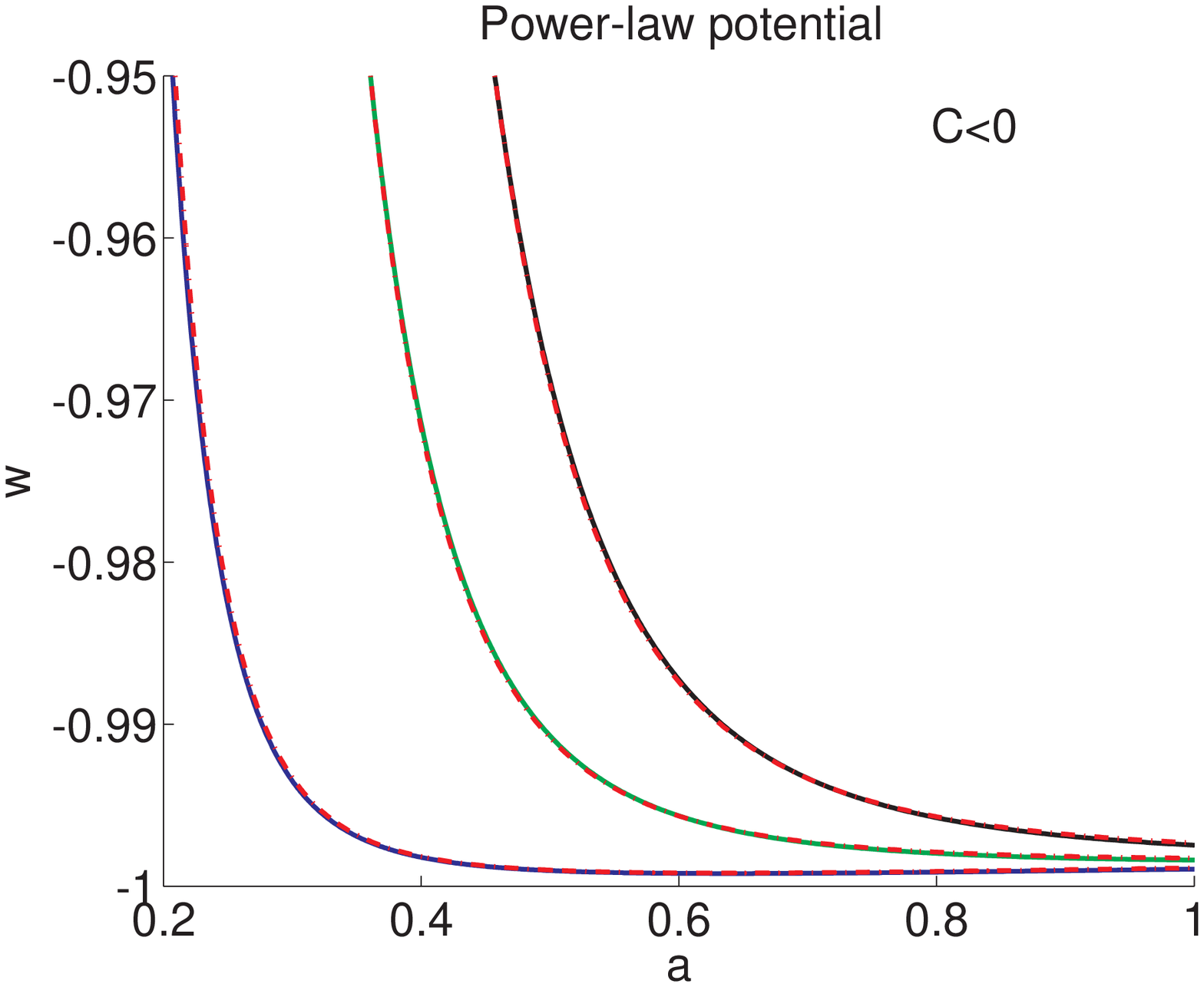,width=7 cm}&\epsfig{file=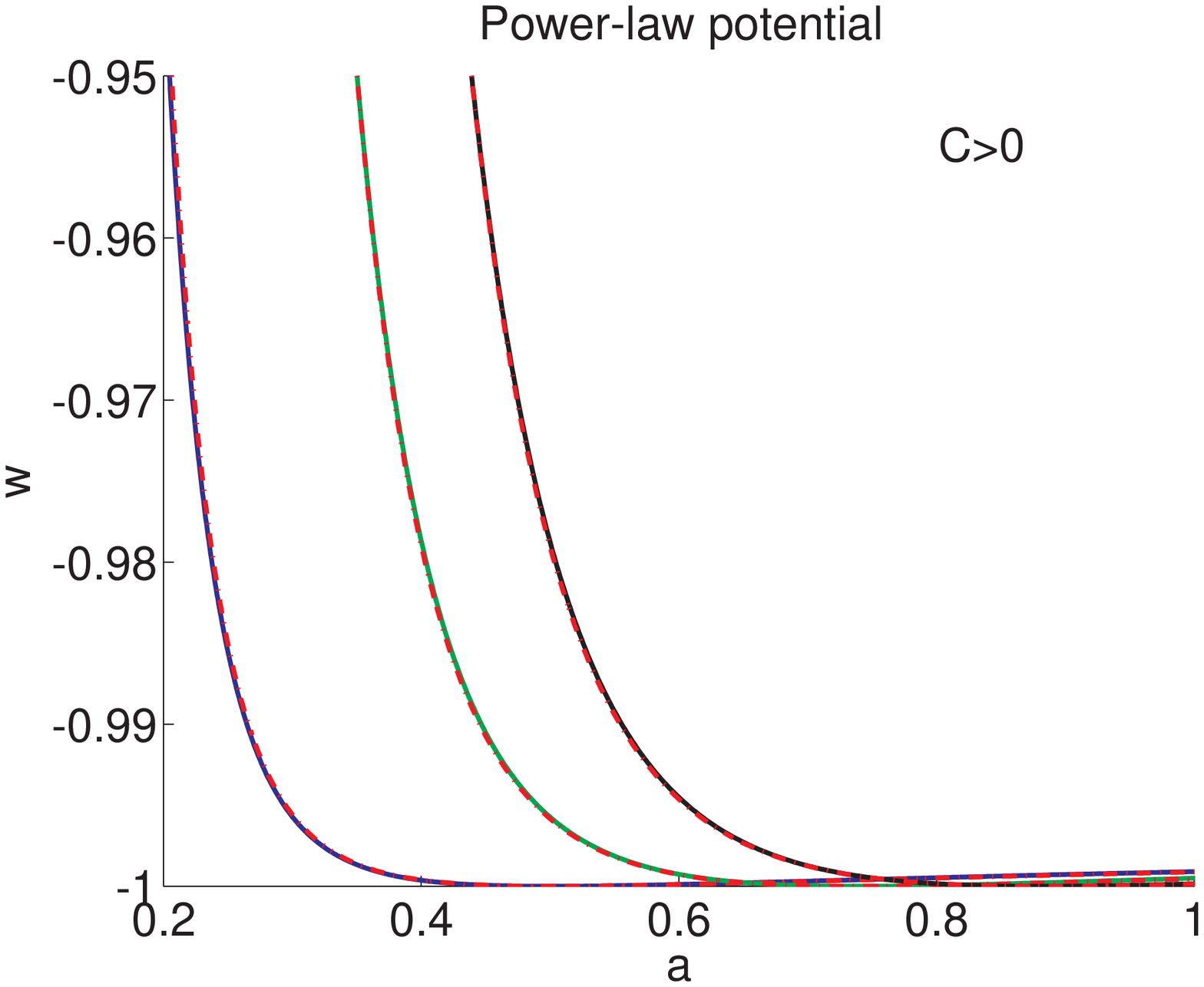,width=7 cm}
\end{tabular}
\caption{\label{w_a_powerlaw}
As \fig{w_a_linear}, for the power-law potential.} 
\end{figure*}
\end{center}

Our results are displayed in Figs. \ref{w_a_linear} and \ref{w_a_powerlaw}.
Note that the agreement between our
analytic approximation and the true (numerically-calculated) evolution is
excellent.
This demonstrates that in the regime where the slow roll conditions hold,
the evolution of $w$ converges to a common set of tracks
irrespective of the details of the underlying quintessence potential. 

In the limit where $\lambda = 0$, these models correspond
to a field
evolving in a flat potential, $V = V_0$; these models have
been dubbed ``skating" models \cite{Linderpaths}.  Since we
are taking $\lambda \ll 1$, it is reasonable to ask
whether the freezing portion of these trajectories can
be well-approximated by a pure skating trajectory.

An approximate skating trajectory will be realized whenever the first term in the numerator
of equation (\ref{exact}) dominates the second term; this corresponds
to the condition
\begin{equation}
\label{sk1}
1+w \gg \frac{\lambda^2}{3} \Omega_\phi.
\end{equation}
Combining this result with equation (\ref{gensol}) (for $C < 0$) gives the following
condition for the evolution to approximate a skating solution:
\begin{equation}
\label{sk2}
-C \gg \tanh^{-1}\sqrt{\Omega_\phi} - \sqrt{\Omega_\phi}.
\end{equation}

Equation (\ref{sk2}) indicates that, for a given $C$, our models approximate a skating
solution at sufficently early times (small $\Omega_\phi$).  We can demonstrate this explicitly by
solving equation (\ref{exact}) for the case $\lambda=0$, giving
\be
\label{wskating}
(1-w)(1+w)\frac{\Omega_\phi^2}{\(1-\Omega_\phi\)^2} = k,
\ee
where $k$ is a constant.  At first glance, it is not obvious how
equations (\ref{wskating}) and (\ref{gensol}) could correspond to the same
behavior for $w(\Omega_\phi$).  However, if we take the limit
of small $\Omega_\phi$ in equation (\ref{gensol}), as indicated by equation
(\ref{sk2}),
and take $1+w \ll 1$ in equation (\ref{wskating}) (since it is only in this
limit that equation (\ref{gensol}) will be valid), we see that
the two expressions are, in fact, equivalent, with $k$ in equation
(\ref{wskating}) given by $k = (2/3)\lambda_0^2 C^2$.

\begin{figure}
\begin{center}
	\epsfig{file=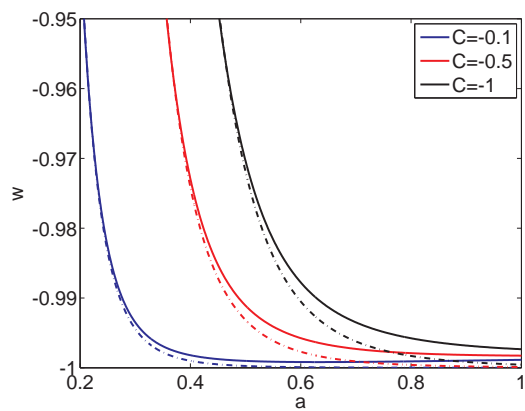,height=55mm}
	\caption{\label{w_skating}
	The equation of state parameter, $w$, as a function of the scale factor,
	$a$, for the linear potential (equation \ref{linearpotential})
	with $\alpha = 0.1$ and the indicated
	values of $C$ (solid curves), along with the corresponding models
	with $\alpha = 0$ (dashed curves).
}
\end{center}
\end{figure}

We explore this equivalence in \fig{w_skating}.  Here we plot several $w(a)$ curves
for the linear potential, along with the evolution of $w$
for the corresponding pure skating potential.
As expected, the two sets of trajectories coincide at early times but
diverge at late times. However, since it is the latter which is the epoch of interest
observationally,
the approximation developed here cannot usefully be replaced by the
solution appropriate to a constant potential.

\section{Discussion}

We have extended earlier work on nearly flat potentials to include
the cases where $\dot \phi \ne 0$ initially.  In these cases, the single
trajectory for $(1+w)/\lambda_0^2$ as a function of $\Omega_\phi$, derived
in Ref. \cite{ScherrerSen1}, becomes a family of trajectories parametrized
by a constant $C$ that depends on the initial value of $\dot \phi$.
All such trajectories show roughly the same behavior:  an initial period
of freezing behavior, for which $w$ decreases with time, followed by
a thawing period, for which $w$ increases.  However, this thawing period can
take place in the future, so these models allow for purely freezing behavior
up to the present.  Note that behavior of this type (with freezing followed
by thawing) was also noted in the Monte Carlo simulations of
Huterer and Peiris \cite{HP}, although
their space of sampled models differs significantly from the class of
models explored here.

Our results differ from the earlier work in Refs. \cite{Watson}
and \cite{Chiba}, who also examined freezing models.  However, these papers
assumed a model that began on a tracking trajectory, with $w$ initially constant
and far
from $-1$, and then examined the freezing evolution as the scalar field began
to influence the expansion.  The models discussed here assume that the potential
is sufficiently flat that $w$
is already close to $-1$ at late times, so we expect very different
evolution between these two cases.

The results presented here can be easily generalized, as in Ref.
\cite{ScherrerSen2}, to phantom models with a negative kinetic term.  The
result is simply equation (\ref{gensol}) multiplied by $-1$ on the right-hand
side.

Obviously, our results apply only to a special set of
quintessence potentials.  However, they do provide an interesting set of
restricted evolutionary paths for such potentials, and potentials satisfying
the slow-roll conditions are a ``natural" way to produce the observed value
of $w$ close to $-1$ today.  One might argue, in that regard, that the
freezing models discussed here are less natural than the thawing
models discussed in Ref. \cite{ScherrerSen1}.  This is a valid criticism,
since in the former case one must tune the initial value of $\dot \phi$
to give $w$ sufficiently close to $-1$ at present.
Furthermore, the evolution discussed here cannot be extrapolated
arbitrarily far into the past, since this would result in an
early universe dominated by scalar field kinetic energy, in contradiction to observations \cite{stiff}.
Thus, the slow roll conditions on $V(\phi)$ cannot be satisfied at arbitrarily
early times; the field must have evolved from a region with a different
form for $V(\phi)$, giving a different evolution for $w(a)$ in the early
universe.
This means that freezing slow-roll models
are necessarily more complicated than the thawing models considered in
Ref. \cite{ScherrerSen1}, which do not have similar problems with their
evolution at early times.
However, in the absence
of any compelling {\it a priori} models for the scalar field, it is
worthwhile to consider all reasonable possibilities.

\section{Acknowledgments}
R.J.S. was supported in part by the Department of Energy
(DE-FG05-85ER40226).


\end{document}